\documentstyle[twocolumn,prd,aps]{revtex}

\begin{document}

\input{feynman}

\bibliographystyle{unsrt}

\author{J.W.~Bos$^1$, D. Chang$^{2,1}$, S.C.~Lee$^1$, Y.C.~Lin, and
H.H.~Shih$^3$}
\address{${}^1$Institute of Physics, Academia Sinica, Taipei, Taiwan
\\${}^2$Department of Physics, National Tsing Hua University, Hsinchu,
Taiwan\\${}^3$Department of Physics and Astronomy, National Central
University, Chungli, Taiwan}
\title{Hyperon weak radiative decays in chiral perturbation theory}

\maketitle

\begin{abstract}
We investigate the leading-order amplitudes for weak radiative decays
of hyperons in chiral perturbation theory. We consistently include
contributions from the next-to-leading order weak-interaction
Lagrangian.  It is shown that due to these terms Hara's theorem is
violated.  The data for the decays of charged hyperons can be easily
accounted for.  However, at this order in the chiral expansion, the
four amplitudes for the decays of neutral hyperons satisfy relations
which are in disagreement with the data.  The asymmetry parameters for
all the decays can not be accounted for without higher-order terms.
We shortly comment on the effect of the 27-plet part of the weak
interaction.
\end{abstract}

\section{Introduction}\label{introduction}

Weak radiative decays of hyperons, $B_i\rightarrow B_f + \gamma$, have
received attention for a long time, both experimentally
\cite{lam,xinlam,xinsin,timm,sip,xim} and theoretically
\cite{hara,gaillard,disp,pole,disp-quark-pole,pole-quark,%
vasa,kamal,quark,vector,qcd-sum-rule,other,neuf,jenk}.  There have
been several theoretical approaches to this problem.  One of the two
major approaches uses hadronic degrees of freedom
\cite{disp,pole,disp-quark-pole,pole-quark} while the alternative is
solely based on the quark picture of hyperons \cite{vasa,kamal,quark}.
While the data are known to be consistent with the lower bounds on the
amplitudes derived from unitarity constraints
\cite{disp,disp-quark-pole}, none of the theoretical models have
managed to give a satisfactorily account on details of the data, in
particular, the rates for the four neutral decay modes.

One of the issues that has been emphasized in the literature is the
apparent violation of Hara's theorem\cite{hara,gaillard}, which states
that the parity-violating amplitudes for $\Sigma^+\rightarrow p+\gamma$
and $\Xi^-\rightarrow \Sigma^- +\gamma$ vanish in the limit of
$SU(3)$ symmetry.  It predicts, in contradiction with experiments,
that the asymmetry parameter for $\Sigma^+\rightarrow p+\gamma$ should
be quite small. (See Ref.\cite{lach} for a review of the relevant
arguments.)

The hadronic models did not have a great deal of success in explaining
the details of the data.  All the models of this type (except that
which include vector mesons) preserve Hara's theorem in their
formulations.  General analysis which include $SU(3)$
breaking\cite{vasa} actually predicts a small and positive asymmetry
for the $\Sigma^+$ decay while the data shows that it is negative and
relatively large.  Models that assume vector-meson dominance
\cite{vector} can introduce effects that violate Hara's theorem due to
mixing of the vector meson with the photon.  In models using quarks,
it was pointed out \cite{kamal} that the diagrams in which a $W$-boson
is exchanged between two constituent quarks can give rise to violation
of Hara's theorem.  In addition, models which include vector-meson
dominance are in better agreement with the data, though the situation
is still not satisfactory.  The experimentally observed negative
asymmetry parameter for $\Sigma^+$ decay is best accounted for using
QCD sum-rules \cite{qcd-sum-rule}.  Other approaches can be found in
Refs.~\cite{other,neuf,jenk}.  A detailed overview on both
experimental and theoretical aspects of weak radiative decays of
hyperons is given in Refs.~\cite{timm,lach,review}.

Chiral perturbation theory (ChPT) \cite{cpt,weinberg} has been shown to
be a useful way of describing low energies hadronic processes,
especially those that involve only mesons.  It is an effective field
theory in terms of hadronic degrees of freedom, based on the symmetry
properties of QCD.  For application to processes involving baryons, it
is most consistently formulated in the heavy-baryon formulation
\cite{hbcpt}, in which the $SU(3)$ invariant baryon mass, $\dot{m}$, is
removed by a field transformation (see also Ref.~\cite{weinberg2},
where a similar transformation is carried out).  In this approach an
amplitude for a given process is expanded in external pion
four-momenta, $q$, baryon {\em residual\/} four momenta, $k$, and the
quark mass, $m_s$.  We will neglect the up and down quark mass.  We
will collectively write down $q$, $k$, and $m_s$ as $E$. (As we will
discuss later, we will adopt the convention that $k$ and $m_s$ are of
the same order in the chiral expansion.) The perturbation theory is
reliable only when $E$ is smaller than the chiral symmetry breaking
scale $\Lambda_{\chi}$.  In the heavy-baryon formulation there is an
additional expansion in $1/\dot{m}$.  However, all these terms can be
effectively absorbed in counterterms of the theory \cite{we}.

Weak radiative decays of hyperons have been studied before in the
context of ChPT by Jenkins, Luke, Manohar and Savage \cite{jenk} and
Neufeld \cite{neuf}.  Jenkins {\it et al.} and Neufeld calculated the
amplitude up to the one-loop level.  These loop diagrams give
contributions to the amplitudes which are at least of order $O(E^2)$
in the chiral expansion.  However, tree-level direct emission diagrams
from the next-to-leading order weak Lagrangian \cite{we}, which give
contribution of order $O(E)$ to the amplitudes, were not considered.
The reasons such terms might be neglected consistently is the fact
that they are not needed for renormalization.  However, in general,
since ChPT should be based on a most general Lagrangian
\cite{weinberg}, they should be included also.  We will see that as a
consequence of not taking these terms into account, the analysis for
weak radiative decay of both Jenkins {\it et al.} and Neufeld should
satisfy Hara's theorem.

In this paper we consistently calculate the leading-order amplitude of
weak radiative decays of hyperons in ChPT.  At this order, no loop
contribution need to be considered. However, one does need to take into
account the higher-order terms in the weak chiral Lagrangian.  We will
show that it gives rise to violation of Hara's theorem.  As a
consequence the decay rates for the charged decays $\Sigma^+\rightarrow
p+\gamma$ and $\Xi^-\rightarrow\Sigma^-+\gamma$ can be accounted for
consistently.  We will show that, in leading order, ChPT predicts the
ratios of the decay amplitudes of all the neutral channels as functions
of the baryon masses only.  We will compare these predictions with the
data.  Furthermore, the asymmetry parameters still vanish in this
leading order calculation.  We shall explain why this is not
necessarily inconsistent with the data in the expansion scheme of
ChPT.  We will also discuss the contribution of the 27-plet component
of the weak Lagrangian to the amplitude.

This paper is organized as follows. We will start in the next section
by discussing the general formalism of hyperon weak radiative
decay. In Sec.~\ref{amplitude} we will calculate the amplitude in
leading order ChPT, including contributions from the weak and strong
interaction Lagrangian with higher order terms.  In Sec.~\ref{27} we
will discuss briefly the contribution from the 27-plet.  Next, in
Sec.~\ref{discussion}, the parameters appearing in the expressions are
discussed and compared with data.  Finally, Sec.~\ref{summary}
contains a summary and our conclusions.

\section{General formalism}

In this section we will consider some general features of weak
radiative hyperon decay, and summarize the data.  To deal with
baryons, we will work in the heavy-baryon formalism \cite{hbcpt},
briefly outlined in the appendix.

As shown in the appendix, in the heavy-baryon formalism and in the
gauge
\[
	v\cdot A = 0\;,
\]
the general amplitude for the hyperon weak radiative decay process
\begin{equation}
	 B_i (m_i) \rightarrow B_f (m_f)+\gamma
\end{equation}
is given by
\begin{eqnarray}
\label{gen_ampl}
	\epsilon_\mu(q)&&{\cal M}^\mu = \nonumber\\
	&&2\epsilon_\mu(q)\bar{U}_{v}(k')\bigl(q_\nu[S_v^\mu,S_v^\nu]A+
	\Delta m S_v^\mu B\bigr)U_{v}(k)\;,
\end{eqnarray}
where $U_v$ and $\bar{U}_v$ are the heavy-baryon spinors of the
initial and final baryons, respectively, and $\epsilon_\mu$ is the
photon polarization vector.  The momenta are defined in Fig.~1.  The
``form factors'' $A$ and $B$ in Eq.~(\ref{gen_ampl}) correspond to the
parity-conserving and parity-violating part of the amplitude,
respectively.  The factor $\Delta m\equiv m_{i}-m_{f}$ multiplying $B$
appears by convention: it is introduced in order to reproduce the
parity-violating form factor in the conventional relativistic
formalism (see Eq.~(\ref{rel}) of the appendix).  The factor plays a
crucial role in the discussion of Hara's theorem.

Hara's theorem concerns the parity-violating amplitude in the limit of
$U$-spin symmetry. ($U$-spin transformations interchange an $s$ and $d$
quark.) Assuming $U$-spin symmetry Hara's theorem can be easily
obtained from Eq.~(\ref{gen_ampl}). If we have $U$-spin symmetry the
mass difference between $p$ and $\Sigma$ vanishes:
\begin{equation}
	m_{\Sigma^+}-m_p=0\;.
\end{equation} 
If we also assume that the parity-violating form factor $B$ has {\em no\/}
pole in $m_{\Sigma^+}-m_p$, we find from Eq.~(\ref{gen_ampl}) that the
parity violating amplitude for $\Sigma^+ \rightarrow p + \gamma$
vanishes.  However, as we will see in the following, the assumption
that $B$ is non-singular may not be correct in the framework of ChPT.

There are two possible independent observables in this process.  Using
Eq.~(\ref{gen_ampl}) and the photon-polarization sum in the gauge
$v\cdot A = 0$,
\begin{equation}
	\sum_{\lambda}\epsilon^\mu_\lambda(k)\epsilon^\nu_\lambda(k)=
	-g^{\mu\nu}-\frac{k^\mu k^\nu}{(v\cdot k)^2} +
	\frac{k^\mu v^\nu+k^\nu v^\mu}{v\cdot k}\;,
\end{equation}
we find for the decay rate the usual expression
\begin{equation}
\label{rate}
	\Gamma=\frac{\omega^3}{\pi}(|A|^2 + |B|^2)\;,
\end{equation}
where $\omega$ is the photon energy in the lab-frame given by
\begin{equation}
	\omega=\frac{m_i^2-m_f^2}{2m_i}\;.
\end{equation}
As required the decay rates are regular in the chiral limit even if
the form factors are singular, since the aforementioned potential
singular behavior of $A$ and $B$ is compensated by the phase-space
factor $\omega$.  The second observable, related to the angular
dependence, is the asymmetry parameter given by
\begin{equation}
\label{ap}
	\alpha = \frac{2\Re(AB^\ast)}{|A|^2+|B|^2}\;.
\end{equation}
The present data on the decay rates and asymmetry parameters is
summarized in Table~\ref{data}.

\section{Leading order amplitude}\label{amplitude}

\subsection{Lagrangian and Feynman diagrams}

We will now turn to the calculation of the hyperon weak radiative
decays in leading-order ChPT in the heavy-baryon formalism.  The
necessary weak ChPT Lagrangian, up to terms of order $E$, has been
given in Ref.~\cite{we}.  We shall consider only the $CP$ even part of
the Lagrangian.  The diagrams contributing to the leading-order
amplitude are tree diagrams given in Fig.~2.  There are two kinds of
diagrams: the direct emission diagrams Fig.~2a, and the baryon
pole-diagrams Fig.~2b.  Loop diagrams can be omitted in our
calculation, since they give rise to contributions of higher order.

Since the full Lagrangian, including the Lagrangian in the
weak-interaction sector, has been given elsewhere, we give here only
the terms directly relevant to the hyperon weak radiative decay in
leading order.  The baryons are represented by the usual $SU(3)$
matrix
\begin{equation}
	H=\left(\begin{array}{ccc}\frac{1}{\sqrt{6}}\Lambda+
	\frac{1}{\sqrt{2}}\Sigma^0 &
	\Sigma^+&p\\\Sigma^-&
	\frac{1}{\sqrt{6}}\Lambda-\frac{1}{\sqrt{2}}\Sigma^0 &n\\
	\Xi^-&\Xi^0&-\frac{2}{\sqrt{6}}\Lambda\end{array}\right) \;,
\end{equation}
and, since pions do not enter this tree-level description, we can take
for the other fields the expansions
\begin{eqnarray}
	&&D^\mu=\partial^\mu-ieQA^\mu\;,\;\;
	\Delta^\mu=0\;,\;\;\sigma=\chi
\nonumber
\\
	&&\rho=0\;,\;\;\lambda=\lambda_6,\;\;\lambda'=\lambda_7\;,
\end{eqnarray}
where $Q$ is the quark charge-matrix
\begin{equation}
\label{charge}
	Q=\frac{1}{3}\,\text{diag}(2,-1,-1)\;,
\end{equation}
$\lambda_{6,7}$ are the Gell-Mann matrices (giving rise to $|\Delta S|
=1$ transitions),
\begin{equation}
\lambda_6=\left(\begin{array}{ccc}
 	0&0&0\\0&0&1\\0&1&0\end{array}\right)\;,\;\;
\lambda_7=\left(\begin{array}{ccc}
 	0&0&0\\0&0&-i\\0&i&0\end{array}\right)\;,
\end{equation}
and $\chi$ the $SU(3)$-breaking mass matrix
\begin{equation}
	m_s\,\text{diag}(0,0,1)\;.
\end{equation}
In leading order, the the decuplet does not play a role and we can
restrict ourselves to the terms
\begin{eqnarray}
	{\cal L}_s^{(1,0)}&=&i\text{Tr}\Bigl[\,\bar{H}[v\cdot D,H]\,
	\Bigr]\;,
\\
	{\cal L}_s^{(0,1)}&=&A_{1}\text{Tr}\Bigl[\,\bar{H}
	\{\sigma,H\}\,\Bigr]+A_{2}\text{Tr}\Bigl[\,\bar{H}
	[\sigma,H]\,\Bigr]
\nonumber
\\
	&&\mbox{}+A_3\text{Tr}\Bigl[\,\bar{H}H\,\Bigr]
	\times\text{Tr}\Bigl[\,\sigma\,\Bigr]\;,
\label{mass}
\\
	{\cal L}_s^{(2,0)}&=&B_1\text{Tr}\Bigl[\,\bar{H}[D^\mu,
	[D_\mu,H]]\,\Bigr]
\nonumber
\\
	&&\mbox{}+B_3\text{Tr}\Bigl[\,\bar{H}[S_v^\mu,S_v^\nu]
	\{[D_\mu,D_\nu],H\}\,\Bigr]
\nonumber
\\
	&&\mbox{}+
	B_4\text{Tr}\Bigl[\,\bar{H}[S_v^\mu,S_v^\nu]
	[[D_\mu,D_\nu],H]\,\Bigr]\;,
\label{mm}
\\
	{\cal L}_w^{(0,0)}&=&h_D\text{Tr}\Bigl[\,\bar{H}\{\lambda,H\}
	\,\Bigr]
	+h_F\text{Tr}\Bigl[\,\bar{H}[\lambda,H] \,\Bigr]\;,
\\
 	{\cal L}_w^{(1,0)}&=&ia_{5}\text{Tr}\Bigl[\,\bar{H}S_v^\mu
	\{\lambda,[D_\mu,H]\}+\bar{H}S_v^\mu[D_\mu,\{\lambda,H\}]\,
	\Bigr]
\nonumber
\\
	&&\mbox{}+ia_{6}\text{Tr}\Bigl[\,\bar{H}S_v^\mu
	[\lambda,[D_\mu,H]]+\bar{H}
	S_v^\mu[D_\mu,[\lambda,H]]\,\Bigr]
\nonumber
\\
	&&\mbox{}+a_{7}\text{Tr}\Bigl[\,\bar{H}S^\mu_v\{[D_\mu,
	\lambda'],H\}\,\Bigr]
\nonumber
\\
	&&\mbox{}+a_{8}\text{Tr}\Bigl[\,\bar{H}S^\mu_v[[D_\mu,
	\lambda'],H]\,\Bigr]\;.
\label{weak}
\end{eqnarray}
In our notation the superscript $(i,j)$ denotes a term of order
$k^im_s^j$ in the chiral expansion, and the subscripts $s$ and $w$
identify the strong and weak interaction, respectively.  As mentioned
in Sec.~(\ref{introduction}), the terms in ${\cal L}_w^{(1,0)}$,
Eq.~(\ref{weak}), were not taken into account in previous ChPT
calculations.

The parameters $A_1, A_2$ and $A_3$ in the strong Lagrangian ${\cal
L}_s^{(0,1)}$ determine the four $SU(2)$ invariant masses of the octet
baryons up to first order in $m_s$.  Therefore, they provide one
prediction, which is the Gell-Mann--Okubo mass relation, and fit the
physical baryon masses within about 5 \%.  We will choose these
parameters such that the baryon masses are fitted best.  Except for
these mass terms in the strong Lagrangian ${\cal L}_s^{(0,1)}$, all
the terms in the Lagrangian obey $SU(3)$ symmetry.  Therefore, all
$SU(3)$-breaking effects in the amplitudes in our formulation are due
to the intermediate baryon propagator in the pole-diagrams in
Fig.~(2b).  Since we have chosen $m_u = m_d = 0$, and the small mass
effects due to the electromagnetic interaction can be ignored here,
the baryon masses obey isospin symmetry and we will use the obvious
notation $m_N$, $m_\Lambda$, $m_{\Sigma}$, and $m_{\Xi}$ to represent
the average mass of the corresponding isospin multiplets.  Note that
in calculating the decay rates the phase space gives rise to
additional sources of $SU(3)$ breaking mass differences.  However, in
that case, we shall use the observed masses.

The terms with $B_1, \ldots, B_3$ in the strong-interaction Lagrangian
${\cal L}_s^{(2,0)}$ enter the amplitude for weak radiative decays
through the baryon electromagnetic vertex in the pole-diagrams of
Fig.~(2b), while $h_D$ and $h_F$ in the weak-interaction Lagrangian
${\cal L}_w^{(0,0)}$ enter through the weak baryon-mixing in these
same diagrams.

Finally, the terms containing the parameters $a_5, \ldots, a_8$ in the
weak Lagrangian ${\cal L}^{(1,0)}_w$ give rise to the direct emission
diagrams shown in Fig.~(2a).  However, since $[\lambda_7 , Q] = 0$ the
terms with $a_7$ and $a_8$ do not contribute to hyperon weak radiative
decays in leading order, and can be ignored in the following.

The diagrams in Fig.~2 lead to the following results for the
parity-conserving form factors A in leading order ChPT,
\begin{mathletters}
\label{A}
\begin{equation}
A_{\Lambda\rightarrow n\gamma}=-\frac{eB_3}{3\sqrt{6}}\left(
\frac{3h_F+h_D}{m_\Lambda-m_N}-3\frac{h_F-h_D}{m_\Sigma
-m_N}\right)\;,
\end{equation}
\begin{equation}A_{\Sigma^+\rightarrow p\gamma}=0\;,
\end{equation}
\begin{equation}A_{\Sigma^0\rightarrow n\gamma}=\sqrt{3}A_{\Lambda\rightarrow 
n\gamma}\;,
\end{equation}
\begin{eqnarray}
\label{d}
A_{\Xi^0\rightarrow\Lambda\gamma}&=&-\frac{eB_3}{3\sqrt{6}}\left(
\frac{3h_F-h_D}{m_\Xi-m_\Lambda}-3\frac{h_F+h_D}{m_\Xi-m_\Sigma}\right)
\nonumber\\
&=&
-\frac{m_\Lambda-m_N}{m_{\Xi}-m_\Lambda}
\frac{m_\Sigma-m_N}{m_\Xi-m_\Sigma}
A_{\Lambda\rightarrow n\gamma}\;,
\end{eqnarray}
\begin{eqnarray}
A_{\Xi^0\rightarrow\Sigma^0\gamma}
&=&\sqrt{3}A_{\Xi^0\rightarrow\Lambda\gamma}
\nonumber\\
&=&-\sqrt{3}\frac{m_\Lambda-m_N}{m_{\Xi}-m_\Lambda}
\frac{m_\Sigma-m_N}{m_\Xi-m_\Sigma}
A_{\Lambda\rightarrow n\gamma}\;,
\end{eqnarray}
and
\begin{equation}
A_{\Xi^-\rightarrow\Sigma^-\gamma}=0\;.
\end{equation}
\end{mathletters}
The final result in Eq.~(\ref{d}) follows from the Gell-Mann--Okubo
mass relation which is, as discussed above, satisfied to the order we
consider.  The expressions for the parity-conserving amplitude was also
be obtained \cite{gaillard} from arguments based on the $U$-spin
symmetry (which is satisfied in this order due to the structure of the
quark charge-matrix $Q$ in Eq.~(\ref{charge})) of the magnetic
moments. 

For the parity-violating amplitude $B$ we find,
\begin{mathletters}
\label{B}
\begin{equation}
B_{\Lambda\rightarrow n\gamma}=0\;,
\end{equation}
\begin{equation}
B_{\Sigma^+\rightarrow p\gamma}=\frac{e(a_5-a_6)}
{m_\Sigma-m_N}\;,
\end{equation}
\begin{equation}
B_{\Sigma^0\rightarrow n\gamma}=0\;,
\end{equation}
\begin{equation}
B_{\Xi^0\rightarrow\Lambda\gamma}=0\;,
\end{equation}
\begin{equation}
B_{\Xi^0\rightarrow\Sigma^0\gamma}=0\;,
\end{equation}
and
\begin{equation}
B_{\Xi^-\rightarrow\Sigma^-\gamma}=-\frac{e(a_5+a_6)}{m_\Xi-
m_\Sigma}\;.
\end{equation}
\end{mathletters}
Note that the mass differences, $\Delta m$, in the denominators in
Eq.~(\ref{B}) arise because of the $\Delta m$ in the parity-violating
part of the general amplitude Eq.~(\ref{gen_ampl}).  From the point of
view of ChPT, $a_i$ are the fundamental parameters that should be
treated as constants, i.e., they can not compensate for the $\Delta m$
in the denominators in Eq.~(\ref{B}).  As a result, the $B$ form
factors become singular in the $SU(3)$ invariant limit, contrary to
the usual implicit assumption in the derivation of Hara's theorem.

Let us now take a closer look at our results.  The pole-diagrams only
contribute to the parity-conserving form factor A, in accordance with
the Lee-Swift theorem \cite{lee-swift}. Since the pole contributions
to the charged decay modes $\Sigma^+\rightarrow p+\gamma$ and
$\Xi^-\rightarrow\Sigma^-+\gamma$ cancel, we find a nonzero
parity-conserving form factor only for the neutral decays $\Lambda
\rightarrow n+\gamma$, $\Sigma^0\rightarrow n+\gamma$,
$\Xi^0\rightarrow \Lambda+\gamma$, and $\Xi^0\rightarrow
\Sigma^0+\gamma$.

The terms in weak Lagrangian in Eq.~(\ref{weak}) all contain $[Q,H]$.
Since $[Q,H]=0$ for neutral baryons, the direct emission diagrams do
not contribute to any of the neutral decays.  For the same reason also
the parameters $B_1$ and $B_4$ do not give a contributions to the
neutral decays in the pole diagrams.

Since for all the decays either $A$ or $B$ vanishes we immediately
conclude that the asymmetry parameters, defined by Eq.~(\ref{ap}),
still vanishes in this (leading) order.  However, we can show, by
qualitative arguments, that this does not need to imply a
contradiction between ChPT and the measured asymmetry parameters in
Table~\ref{data}. Assume, in the spirit of ChPT, that the charged
decay modes can be expanded as
\begin{equation}
	A = a_1 \lambda\;;\;\; B = b_0 + \lambda b_1,
\end{equation}
with $\lambda\approx 0.3\; (\approx m_s/\Lambda_{\chi})$, and that
$a_1$, $b_0$, and $b_1$ are of about equal magnitude. It leads, using
Eq.~(\ref{ap}), to an asymmetry parameter of about $0.6$ in magnitude,
which is roughly in agreement with the data.

In leading order, we find from Eq.~(\ref{A})
\begin{equation}
	{\cal M}^\mu_{\Sigma^0\rightarrow n+\gamma}=\sqrt{3}
	{\cal M}^\mu_{\Lambda\rightarrow n+\gamma}\;
\end{equation}
\begin{equation}
	{\cal M}^\mu_{\Xi^0\rightarrow\Lambda+\gamma}=-
	\frac{1}{\sqrt{3}}
	\frac{m_\Lambda-m_N}{m_{\Xi}-m_\Lambda}
	\frac{m_\Sigma-m_N}{m_\Xi-m_\Sigma}
	{\cal M}^\mu_{\Sigma^0\rightarrow n+\gamma}\;
\end{equation}
\begin{equation}
\label{relations}
	{\cal M}^\mu_{\Xi^0\rightarrow\Sigma^0+\gamma}=\sqrt{3}
	{\cal M}^\mu_{\Xi^0\rightarrow\Lambda+\gamma} \;.
\end{equation}
Therefore, all ratios of the neutral decay amplitudes depend only on
the baryon masses and not on any constant from ${\cal L}^{(2,0)}_s$,
${\cal L}^{(0,0)}_w$ or ${\cal L}^{(1,0)}_w$.  Their magnitudes, on the
other hand, also depend on a particular combination of the three
parameters $B_3$, $h_D$, $h_F$.

\section{Inclusion of the 27-plet}
\label{27}

The Lagrangian Eq.~(\ref{weak}) corresponds to the part of the weak
interaction that transforms as $(8_L,1_R)$ under $SU(3)_L \times
SU(3)_R$.  Since the weak interaction consists of a product of two
left-handed flavor-SU(3) currents, the effective chiral Lagrangian
also has a $(27_L,1_R)$ component.  This 27-plet part of the weak
Lagrangian can be included in the same way as the octet part
\cite{we}.  Closer inspection shows that inclusion of the 27-plet
corresponds simply to replacing $h_D$ and $h_F$ in Eqs.~(\ref{A})
and~(\ref{B}) by
\begin{equation}
	h_D \rightarrow h_D + 2 h_{27}
\end{equation}
in the charged channels,
and by
\begin{equation}
	h_D \rightarrow h_D - 3 h_{27}
\end{equation}
in the neutral channels, where $h_{27}$ is the coupling constant from
the leading-order weak interaction Lagrangian that transforms as a
27-plet.  From the $\Delta I = 1/2$ rule of the weak nonleptonic
decays, $h_{27}$ is expected to be a small parameter compared to $h_D$
and $h_F$.  While it enters the charged and neutral channels
differently, the inclusion of the 27-plet into the analysis does not
alter our results, both for the decay rates and the asymmetry
parameters.

\section{Discussion of the results}\label{discussion}

Before discussing the relations Eq.~(\ref{relations}) we first
investigate closer the different parameters in Eqs.~(\ref{A})
and~(\ref{B}) which contribute to the amplitude. Since we find a
distinct results for the neutral and charged channels, we will discuss
them separately.

\subsection{Charged channels and Hara's theorem}\label{charged}

Relevant for the two charged channels are the two parameters $a_5$ and
$a_6$ from the next-to-leading order weak Lagrangian ${\cal
L}_w^{(1,0)}$.  Clearly, we can fit these two parameters to the two
experimental decay rates as given in Table~\ref{data}.  Due to the
quadratic relation between the amplitude and the decay rate we find
the following four combinations for the (dimensionless) parameters
$a_5$ and $a_6$:
\begin{equation}
\label{a5a6}
	a_5=2.8\epsilon\times 10^{-8}\text{ and }a_6=-1.6\epsilon
	\times 10^{-8}\;,
\end{equation}
or
\begin{equation}
	a_5=-1.6\epsilon\times10^{-8}\text{ and }a_6=2.8\epsilon
	\times 10^{-8}\;,
\end{equation}
where $\epsilon = \pm 1$. With these parameters the experimental data
is reproduced. However, no prediction can be made yet.

More interestingly, due to the direct emission diagrams from ${\cal
L}^{(1,0)}_w$, Hara's theorem is violated: Even in the $U$-spin
symmetric limit, the parity-violating part of the decay amplitude is
nonzero.

\subsection{Neutral channels}

As can be seen in Eq.~(\ref{A}) the relevant parameters for the
neutral channels are $B_3$, $h_D$, and $h_F$.  The parameter $B_3$ is
from the strong interaction Lagrangian Eq.~(\ref{mm}), and determines
(together with the constant $B_4$) the magnetic moments of the octet
baryons.  The first column of Table~\ref{t1} shows the magnetic
moments of the baryons in leading order ChPT.  Fitting $B_3$ to the
experimental data, shown in the second column of Table~\ref{t1}, gives
\begin{equation}
\label{b3}
	B_3 = -1.3 \text{ GeV}^{-1}\;.
\end{equation}
The resulting fitted magnetic moments as given in the third column
of Table~\ref{t1}. Note that these results give rise to the Coleman and 
Glashow \cite{cole} relations between the baryon magnetic moments.

The parameters $h_D$ and $h_F$ can be obtained from hyperon
nonleptonic decay.  Considering $s$-wave nonleptonic decay data gives
the best values \cite{jenk92.2},
\begin{equation}
\label{a}
	h_D=-0.58\mu\;;\;\; h_F=1.40\mu\;,
\end{equation}
where $\mu = G_Fm_{\pi}^2\sqrt{2}f_\pi\approx 3 \times 10^{-8}$ GeV.

Using these values, together with that for $B_3$, we arrive at the
decay rates for hyperon radiative decay as given in Table~\ref{fit}.
It shows a huge disagreement with the observed rates. The difference
between the experimental and predicted decay rate is more than a
factor of 200 for $\Lambda\rightarrow n+\gamma$, while the difference
for the other channels is about two orders of magnitude.

This discrepancy, however, is highly dependent on the values used for
$h_D$ and $h_F$.  Independent of any particular values chosen for the
parameters from the Lagrangian we can use Eq.~(\ref{relations}) to
predict three ratios between the four neutral decay rates.  We find
\begin{equation}
	\frac{\Gamma_{\Sigma^0\rightarrow n+\gamma}}
	{\Gamma_{\Lambda\rightarrow n+\gamma}}=8.1\;,
\end{equation}
\begin{equation}
	\frac{\Gamma_{\Xi^0\rightarrow \Lambda+\gamma}}
	{\Gamma_{\Lambda\rightarrow n+\gamma}}=4.9\; (0.59\pm0.14)\;,
\end{equation}
\begin{equation}
	\frac{\Gamma_{\Xi^0\rightarrow \Sigma^0+\gamma}}
	{\Gamma_{\Lambda\rightarrow n+\gamma}} =3.7\; (1.99 \pm 0.42)\;,
\end{equation}
where the experimental values obtained from Table~\ref{data} are
written between parentheses.  While the first ratio can not be obtained
from experimental data, since $\Sigma^0\rightarrow n+\gamma$ has not
been measured, the predicted values for the other ratios is are only
within about a factor 8 and 2 in accordance with the observed ratios.
In contrast to the case of the asymmetry parameters one can not hope
that this disagreement may be resolved by higher order effects, if ChPT
is a consistent expansion scheme with the next-to-leading order
corrections about 30 \% suppressed.

\section{Summary and conclusions}\label{summary}

We have studied the process of hyperon weak radiative decay in the
framework of heavy baryon chiral perturbation theory \cite{hbcpt}.  In
particular we have put emphasis on the effect of including a complete
next-to-leading order weak Lagrangian \cite{we} in the description.
We used it to obtain the leading order decay amplitudes. In previous
calculations of hyperon weak radiative decays these contributions to
the amplitudes have been ignored.

In the leading order, all the ratios for decay amplitudes of the four
{\em neutral channels\/}, $\Lambda\rightarrow n+\gamma$,
$\Sigma^0\rightarrow n+\gamma$, $\Xi^0\rightarrow\Sigma^0+\gamma$, and
$\Xi^0\rightarrow\Lambda+\gamma$, are simple functions of the baryon
masses only.  This follows from the observation that, for the neutral
channels, the direct-emission contribution vanishes and only
pole-diagrams contribute.  Compared with experiment these ratios are
up to a factor 8 off.  It is interesting to note that taking the
values of $h_D$ and $h_F$ from  analysis of nonleptonic weak
decays of hyperons leads to predictions which disagree with the
observed decay rates for more than two orders of magnitude.

Clearly, these disagreements indicate something deficient about the
applications of ChPT to the neutral decays.  One possible solution to
this problem is to assume that at leading order, the weak Lagrangian
consists of additional terms, in particular terms that are not
invariant under $U$-spin transformations, or $CPS$, for some reason.
Alternatively, it may indicate that resonances, such as the vector
mesons, have to included in the analysis of ChPT as hinted by the
relative success of the vector-meson-dominance models \cite{vector}.

For the two {\em charged channels\/} $\Sigma^+\rightarrow p+\gamma$
and $\Xi^-\rightarrow \Sigma^-+\gamma$ no such relations can be
extract from ChPT to this order.  Contrary to the case of the neutral
channels, only the direct emission diagrams contribute.  The
contributions of all the pole-diagrams totally cancel each other at
this order.  The two observed decay rates can be used to fix exactly
the two parameters $a_5$ and $a_6$ from the next-to-leading order weak
Lagrangian.  Although we derive no prediction within the charged
channels, the parameters extracted will be relevant for other
observables such as weak semi-leptonic decays, making a future
comparison feasible.

We have shown that, even in leading order ChPT, Hara's theorem is
violated by direct emission diagrams contributing to the
parity-violating part of the amplitude.  Going through the original
assumptions in the proof of the theorem, our result indicates that the
parity-violating form factors in the amplitude are singular in the
limit of $U$-spin (or $SU(3)$) symmetry in the context of ChPT.  This
singular behavior leads to the failure of Hara's theorem.

The asymmetry parameters vanish in our leading-order calculation.
However, it can still potentially be accounted for by the higher order
effect within the context of ChPT.  In other words, our results
indicate that the asymmetry parameters are sensitive to loop effects
and parameters in the higher-order Lagrangian.

\section*{Acknowledgments}

This work was supported by the National Science Council of the R.O.C.
under Contracts Nos.  NSC85-2112-M-007-032, NSC85-2112-M-007-029, and
NSC85-2112-M-007-0.

\appendix

\section*{General form of weak radiative decay in heavy-baryon chiral
perturbation theory}

In this appendix we briefly review the heavy-baryon formulation of
ChPT, and give the general amplitude of weak radiative decay in this
formulation.

The nucleon mass in the chiral limit, $\dot{m}$, is comparable with
the chiral symmetry breaking scale $\Lambda_{\chi}$. To make a
consistent chiral expansion possible it can be removed by redefining
the baryon field according to \cite{hbcpt}
\begin{equation}
\label{schaap}
	B_v=\text{e}^{i\dot{m} v\cdot x}B \;,
\end{equation}
where $v^\mu$ is the baryon four-velocity satisfying $v^2=1$.  Next,
one defines the projected fields
\begin{equation}
\label{projfields}
	H =  {P}^+_v B_v\;; \;\; 
	h =  {P}^-_v B_v \;,
\end{equation}
where $P_v^+$ and $P_v^-$ are the projection operators
\begin{equation}
	 {P}^{\pm}_v= \frac{1\pm v\!\!\!\slash}{2} \;.
\end{equation}
The minus component field $h$ is suppressed by $1/\dot{m}$
compared to $H$.
It can be easily seen that, in momentum space, derivatives of $H$
produce powers of
\begin{equation}
	k^\mu = p^\mu -\dot{m} v^\mu \;,
\end{equation}
with $p^\mu$ the four-momentum of the baryon, which is (for processes
at low energies) a small quantitiy.  This {\em residual\/} baryon
momentum, $k$, is the effective expansion parameter in this formulation
of baryon ChPT.  Effects of $1/\dot{m}$ can arise through $h$ in higher
order. However, these $1/\dot{m}$ corrections can be absorbed in the
higher-order counterterms of the theory \cite{we}.

The baryon field $H$ satisfies the Dirac equation
\begin{equation}
\label{dirac}
	( v \cdot k - \bar{m} ) H = 0
\end{equation}
where $\bar{m}\equiv m_B - \dot{m}$ is defined as the {\em residual\/}
mass of the baryon.

The general amplitude for the weak radiative decay
\begin{equation}
	B(p) \rightarrow B'(p') + \gamma
\end{equation}
is given by
\begin{eqnarray}
\label{rel}
	\epsilon_\mu(q)&&{\cal M}^\mu= \nonumber\\
	&&\epsilon_\mu(q)\bar{u}(p')
	i\sigma^{\mu\nu}q_\nu ( A + B \gamma_5 )
	u(p)\;.
\end{eqnarray}
Defining the operator $S_v^\mu \equiv 
(1/2)P_v^+ \gamma^\mu\gamma_5 P_v^+$ and using
\begin{equation}
\label{op1}
	P^+_v \gamma_5 P^+_v = 0\;,\;\;
	P^+_v \gamma^\mu P^+_v =  P_v^+ v^\mu \;,
\end{equation}
\begin{equation}
	P^+_v \sigma^{\mu\nu} P^+_v = -2 i [ S^\mu_v,S^\nu_v ] \;,
\end{equation}
\begin{equation}
	P^+_v \sigma^{\mu\nu} \gamma_5 P^+_v = 
	-2 i (v^\mu S_v^\nu - v^\nu S_v^\mu) \;,
\end{equation}
and Eq.~(\ref{dirac}), we find that the general form of weak radiative
decay in the heavy baryon formalism is given by
\begin{eqnarray}
	\epsilon_\mu(q)&&{\cal M}^\mu= \nonumber\\
	&&2\epsilon_\mu(q)\bar{U}_{v}(k')\Bigl(q_\nu[S_v^\mu,S_v^\nu]A
\nonumber\\&&\mbox{}+
	(S_v^\mu \Delta m_B  - v^\mu S_v \cdot q )
	B\Bigr)U_{v}(k)\;,
\end{eqnarray}
where $\Delta m_B$ is the mass difference between the initial and
final baryon.  In the gauge $v\cdot A = 0 $ we then finally arrive at
the form as is used in the main text Eq.~(\ref{gen_ampl}).

%
%
%
 
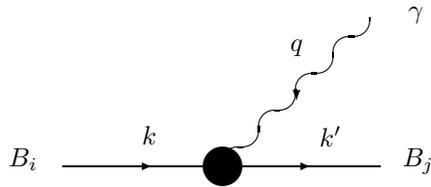
\begin{figure}[t]
\centerline{%
\begin{picture}(15000,10000)
\drawline\fermion[\E\REG](0,2000)[6000]
\put(\pfrontx,\pfronty){\makebox(-1000,0)[br]{$B_i$}}
\drawarrow[\E\ATBASE](\pmidx,\pmidy)
\put(\pmidx,\pmidy){\makebox(500,1500)[tr]{$k$}}
\drawline\fermion[\E\REG](\pbackx,\pbacky)[6000]
\drawarrow[\E\ATBASE](\pmidx,\pmidy)
\put(\pmidx,\pmidy){\makebox(1500,1500)[tr]{$k'$}}
\put(\pbackx,\pbacky){\makebox(2000,0)[br]{$B_j$}}
\put(\pfrontx,\pfronty){\circle*{1500}}
\drawline\photon[\NE\CURLY](\pfrontx,\pfronty)[8]
\put(\pbackx,\pbacky){\makebox(2000,0)[br]{$\gamma$}}
\drawarrow[\S\ATBASE](\pmidx,\pmidy)
\put(\pmidx,\pmidy){\makebox(0,2000)[tc]{$q$}}
\end{picture}}
\caption{Kinematics for weak radiative hyperon decays. $B_i$ and
$B_f$ denote the initial and final hyperon, respectively.
The baryon momenta $k$ and $k'$ are {\em residual\/} momenta,
defined in the appendix.}
\label{fig1}
\end{figure}

\begin{figure}[t]
\centerline{%
\begin{picture}(12000,24000)
\drawline\fermion[\E\REG](2000,18000)[4000]
\drawline\fermion[\E\REG](\pbackx,\pbacky)[4000]
\put(\pfrontx,\pfronty){\makebox(0,0){$\times$}}
\drawline\photon[\NE\CURLY](\pfrontx,\pfronty)[6]
\put(6000,16000){\makebox(0,0){(a)}}
\drawline\fermion[\E\REG](0,9000)[8000]
\put(\pmidx,\pmidy){\makebox(0,0){$\times$}}
\drawline\fermion[\E\REG](\pbackx,\pbacky)[4000]
\drawline\photon[\NE\CURLY](\pfrontx,\pfronty)[6]
\drawline\fermion[\E\REG](0,3000)[4000]
\drawline\fermion[\E\REG](\pbackx,\pbacky)[8000]
\put(\pmidx,\pmidy){\makebox(0,0){$\times$}}
\drawline\photon[\NE\CURLY](\pfrontx,\pfronty)[6]
\put(6000,1000){\makebox(0,0){(b)}}
\end{picture}}
\caption{Feynman diagrams for weak radiative hyperon decay in
leading-order chiral perturbation theory. The cross-sign denotes the
weak interaction.} \end{figure}
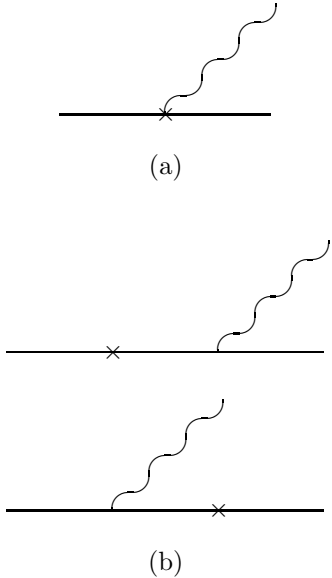

%
%

%
%

\begin{table}[t]
\caption{Present status of decay rates and asymmetry parameters.  The
numbers are the combined weighted mean from Ref.~\protect\cite{lach}.
Both the decay rate and the asymmetry parameter for
$\Sigma^0\rightarrow\Lambda+\gamma$ have not been measured.}
\label{data}
\begin{tabular}{cccc}
$B_i\rightarrow B_f+\gamma$ 
& $\Gamma$ [$10^{-18}$ GeV]
& $\alpha$ 
& Ref.
\\
\hline
$\Lambda\rightarrow n+\gamma$
& $4.07 \pm 0.35$ 
& -- 
& \cite{lam}
\\ 
$\Xi^0\rightarrow\Lambda+\gamma$ 
& $2.4 \pm 0.36$ 
& $0.43 \pm 0.44$ 
& \cite{xinlam}
\\
$\Xi^0\rightarrow\Sigma^0+\gamma$ 
& $8.1 \pm 1.0$ 
& $0.20 \pm 0.32 $ 
& \cite{xinsin}
\\
$\Sigma^+\rightarrow p+\gamma$ 
& $10.1 \pm 0.5$ 
& $-0.76 \pm 0.08 $
& \cite{timm,sip}
\\
$\Xi^-\rightarrow\Sigma^-+\gamma$ 
& $0.51 \pm 0.092$ 
& $1.0 \pm 1.3$ 
& \cite{xim}
\end{tabular}
\end{table}

\begin{table}[t]
\caption{Magnetic moments of the octet baryons and the transitional
moment $\Sigma^0\rightarrow\Lambda+\gamma$ in leading-order chiral
perturbation theory. The first column contains the expressions in
leading-order ChPT, the second column the experimental values, and the
third column the fitted values, with $B_3=-1.13 \text{ GeV}^{-1}$ and
$B_4=-0.82 \text{ GeV}^{-1}$.  The average difference between the
fitted and experimental moments is 19~\%. The constants $B_3$ and
$B_4$ are from the next-to-leading order strong Lagrangian ${\cal
L}_s^{(2,0)}$ (see Eq.~(\protect\ref{mm})). Note that
only the constant $B_3$ plays a role in
hyperon radiative decays.}
\label{t1}
\begin{tabular}{cccc}
$B$ & ${\cal M}_{\text{th}}$ [$e$] & ${\cal M}_{\text{exp}}$ [$\mu_N$]
& ${\cal M}_{\text{fitted}}$ [$\mu_N$] \\ \hline $p$ & $-B_3/3-B_4$ &
2.79 & 2.25 \\ $n$ & $2B_3/3$ & -1.91 & -1.41 \\ $\Lambda$ & $B_3/3$ &
-0.61 & -0.71 \\ $\Sigma^+$ & $-B_3/3-B_4$ & 2.42 & 2.25 \\ $\Sigma^-$
& $-B_3/3+B_4$ & -1.16 & -0.84 \\ $\Xi^-$ & $-B_3/3+B_4$ & -0.68 &
-0.84 \\ $\Xi^0$ & $2B_3/3$ & -1.25 & -1.41 \\
$\Sigma^0\rightarrow\Lambda$ & $B_3/\sqrt{3}$ & $\pm 1.6$ & --
\end{tabular}
\end{table}

\begin{table}[t]
\caption{Decay rates for the four neutral hyperon
radiative decays taking for the parameters $h_D$ and $h_F$ the
values in Eq.~(\protect\ref{a}) obtained from
nonleptonic hyperon decays.
The first
column shows the observed rates from Ref.~\protect\cite{review}.
All decay rates are in units of $10^{-18}$ GeV.}
\label{fit}
\begin{tabular}{ccc}
$B_i\rightarrow B_f+\gamma$ & $\Gamma_{\text{exp}}$  & 
$\Gamma_{ \text{ Eq.~(\protect\ref{a}) } }$ 
\\
\hline
$\Lambda\rightarrow n+\gamma$ & $4.07$ & $0.018$
\\
$\Sigma^0\rightarrow n+\gamma$ & - & $0.16$
\\ 
$\Xi^0\rightarrow\Lambda+\gamma$ & $2.4$ &  $0.087$
\\
$\Xi^0\rightarrow\Sigma^0+\gamma$ & $8.1$ & $0.067$
\end{tabular}
\end{table}

%
%


\begin{thebibliography}{10}

\bibitem{lam}
K.~Larsen {\it et~al.}, Phys. Rev. {\bf D47}, 799 (1993);
S.~F.~Biagi {\it et~al.}, Z. Phys. {\bf C30}, 201 (1986);
B.~L.~Roberts {\it et~al.}, Nucl. Phys. {\bf A479}, 75 (1988);
B.~L.~Roberts {\it et~al.}, Nucl. Phys. (Proc. Suppl.) {\bf B13},
449 (1990);
A.~J.~Noble {\it et~al.}, Phys. Rev. Lett. {\bf 69}, 414 (1992).

\bibitem{xinlam}
C.~James {\it et~al.}, Phys. Rev. Lett. {\bf 64}, 843 (1990);
N.~Yeh {\it et~al.}, Phys. Rev. {\bf D10}, 3545 (1974).

\bibitem{xinsin}
S.~Teige {\it et~al.}, Phys. Rev. Lett. {\bf 63}, 2717 (1989);
J.~R.~Bensinger {\it et~al.}, Phys. Lett. {\bf B215}, 195 (1988).

\bibitem{timm}
S.~Timm {\it et~al.}, Phys. Rev. {\bf D51}, 4638 (1995).

\bibitem{sip}
M.~Foucher {\it et~al.}, Phys. Rev. Lett. {\bf 68}, 3004 (1992);
L.~K.~Gershwin {\it et~al.}, Phys. Rev. {\bf 188}, 2077 (1969);
N.~P.~Hessey {\it et~al.}, Z. Phys. {\bf C42}, 175 (1989);
M.~Kobayashi {\it et~al.}, Phys. Rev. Lett. {\bf 59}, 868 (1987);
S.~F.~Biagi {\it et~al.}, Z. Phys. {\bf C28}, 495 (1985);
A.~Manz {\it et~al.}, Phys. Lett. {\bf B96}, 217 (1980).

\bibitem{xim}
T.~Dubbs {\it et~al.}, Phys. Rev. Lett. {\bf 72}, 808 (1994);
S.~F.~Biagi {\it et~al.}, Z. Phys. {\bf C35}, 143 (1987).

\bibitem{hara}
Y.~Hara, Phys. Rev. Lett. {\bf 12}, 378 (1964);
S.~Y.~Lo, Nuovo Cimento {\bf 37}, 753 (1965);
K.~Tanaka, Phys. Rev. {\bf 140}, B463 (1965);
M.~Gourdin, {\em Unitary symmetry\/} (North-Holland, Amsterdam, 1967).

\bibitem{gaillard}
M.~K.~Gaillard, Nuovo Cimento {\bf A6}, 559 (1971).

\bibitem{disp}
V.~I.~Zakharov and A.~B.~Kaidalov, Sov. J. Nucl. Phys. {\bf 5}, 259
(1967);
G.~R.~Farrar, Phys. Rev. {\bf D4}, 212 (1971);
J.~H.~Reid and N.~N.~Trofimenkoff, Nucl. Phys. {\bf B82}, 397 (1974).

\bibitem{pole}
J.~Pati, Phys. Rev. {\bf 130}, 2097 (1963);
R.~H.~Graham and S.~Pakvasa, Phys. Rev. {\bf 140}, B1144 (1965);
M.~D.~Scadron and L.~Thebaud, Phys. Rev. {\bf D8}, 2190 (1973);
L.~Copley {\it et al.}, Nucl. Phys. {\bf B13}, 303 (1969);
B.~R.~Holstein, Nuovo Cimento {\bf 2A}, 561 (1971);
Y.~I.~Skovpen, Sov. J. Nucl. Phys. {\bf 34}, 773 (1981);
M.~D.~Scadron and M.~Visinescu, Phys. Rev. {\bf D28}, 1117 (1983);
L.~F.~Li and Y.~Liu, Phys. Lett. {\bf B195}, 281 (1987);
M.~K.~Gaillard, Phys. Lett. {\bf B211}, 189 (1988);
G.~Nardulli, Nuovo Cimento {\bf A100}, 485 (1988);
Y.~Liu, Z. Phys. {\bf C45}, 345 (1989);
R.~E.~Karlson and M.~D.~Scadron, Z. Phys. {\bf C52}, 325 (1991);
A.~Maity and P.~Mahato, Nuovo Cimento {\bf A104}, 269 (1991).

\bibitem{disp-quark-pole}
Y.~I.~Kogan and M.~A.~Shifman, Sov. J. Nucl. Phys. {\bf 38}, 628
(1983).

\bibitem{pole-quark}
F.~E.~Close and H.~R.~Rubinstein, Nucl. Phys. {\bf B173}, 477 (1980);
I.~Picek, Phys. Rev. {\bf D21}, 3169 (1980);
M.~B.~Gavela {\it et~al.}, Phys. Lett. {\bf B101}, 417 (1981);
K.~Rauh, Z. Phys. {\bf C10}, 81 (1981);
D.~Palle, Phys. Rev. {\bf D36}, 2863 (1987);
P.~Singer, Phys. Rev. {\bf D42}, 3255 (1990);
R.~E.~Karlsen, W.~H.~Ryan, and M.~D.~Scadron, Phys. Rev. {\bf D43}, 
157 (1991).

\bibitem{vasa}
N.~Vasanti, Phys. Rev. {\bf D13}, 1889 (1976).

\bibitem{kamal}
A.~N.~Kamal and Riazuddin, Phys. Rev. {\bf D28}, 2317 (1983).

\bibitem{quark}
M.~A.~Ahmed and C.~G.~Ross, Phys. Lett. {\bf B59}, 293 (1975);
F.~J.~Gilman and M.~B.~Wise, Phys. Rev. {\bf D19}, 976 (1979);
A.~N.~Kamal and R.~C.~Verma, Phys. Rev. {\bf D26}, 190 (1982);
C.~H.~Lo, Phys. Rev. {\bf D26}, 199 (1982);
S.G.~Kamath, Nucl. Phys. {\bf B198}, 61 (1982);
J.~O.~Eeg, Z. Phys. {\bf C21}, 253 (1984); 
M.~K.~Gaillard, X.~Q.~li and S.~Rudaz, Phys. Lett. {\bf B158}, 158 (1985);
L.~Bergstr\"{o}m and P. Singer, Phys. Lett. {\bf B169}, 297 (1986);
Z. Phys. {\bf C37}, 281 (1988);
R.~Safadi and P.~Singer, Phys. Rev. {\bf D37}, 697 (1988);
R.~Verma and A.~Sharma, Phys. Rev. {\bf D38}, 1443 (1988);
T.~Uppal and R.~Verma, Z. Phys. {\bf C52}, 307 (1991).

\bibitem{vector}
K.~Gavroglu and H.~P.~W.~Gottlieb, Nucl. Phys. {\bf B79}, 168 (1974);
P.~\.Zenczykowski, Phys. Rev. {\bf D40}, 2290 (1989);
{\bf D44}, 1485 (1991);
{\bf D50}, 3285 (1994).

\bibitem{qcd-sum-rule}
V.~M.~Khatsymovski, Sov. J. Nucl. Phys. {\bf 45}, 116 (1987);
{\bf 46}, 768(E) (1987);
{\bf 46}, 496 (1987);
I.~I.~Balitsky, V.~M.~Braun, and A.~V.~Kolesnichenko, Sov. J. Nucl. Phys.
{\bf 44}, 1028 (1986);
Nucl. Phys. {\bf B312}, 509 (1989);
C.~Goldman and C.~O.~Escobar, Phys. Rev. {\bf D40}, 106 (1989).

\bibitem{other}
M.~A.~Shifman, A.~I.~Vainshtein, and V.~I.~Zakharov, Phys. Rev. {\bf D18},
2583 (1978);
A.~Cohen, Phys. Lett. {\bf B160}, 177 (1985);
W.~F.~Kao and H.~J.~Schnitzer, Phys. Rev. {\bf D37}, 1912 (1988);
Phys. Lett. {\bf B183}, 361 (1987);
R.~W.~Brown and E.~A.~Paschos, Nucl. Phys. {\bf B319}, 623 (1989);
P.~Asthana and A.~Kamal, Few Body Systems {\bf 11}, 1 (1991).

\bibitem{neuf}
H.~Neufeld, Nucl. Phys. {\bf {\bf B402}},  166  (1993).

\bibitem{jenk}
E.~Jenkins, M.~Luke, A.~V.~Manohar, and M.~J.~Savage, Nucl. Phys. {\bf {\bf
  B397}},  84  (1993).

\bibitem{lach}
J.~Lach and P.~\.Zenczykowski, Int. J. Mod. Phys. {\bf A10}, 3817 (1995).

\bibitem{review}
B.~Bassalleck, Nucl. Phys. {\bf A547}, 299 (1992).

\bibitem{cpt}
S.~Weinberg, Phys. Rev. Lett. {\bf 18}, 188 (1967);
J.~Schwinger, Phys. Lett. {\bf B24}, 473 (1967);
R.~Dashen, Phys. Rev. {\bf 183}, 1245 (1969);
R.~Dashen and M.~Weinstein, Phys. Rev. {\bf 183}, 1261 (1969);
{\bf 188}, 2330 (1969);
L.~F.~Li and H.~Pagels, Phys. Rev. Lett. {\bf 26}, 1204 (1971);
{\bf 27}, 1089 (1971); Phys. Rev. {\bf D5}, 1509 (1972);
P.~Langacker and H.~Pagels, Phys. Rev. {\bf D8}, 4595 (1973);
{\bf D10}, 2904 (1974);
J.~Gasser and H.~Leutwyler, Phys. Lett. {\bf B125}, 321 (1983);
Ann. Phys. {\bf 158}, 142 (1984);
Nucl. Phys. {\bf B250} (1985) 465, 517, 539.
A.~V.~Manohar and H.~Georgi, Nucl. Phys. {\bf B234}, 189 (1984);
J.~Bijnens, H.~Sonoda, and M.~Wise, Nucl. Phys. {\bf B261}, 185 (1985);
J.~Gasser, M.~Sainio, and A.~\u{S}varc, Nucl. Phys. {\bf B307}, 779
(1988);
A.~Krause, Helv. Phys. Acta {\bf 63}, 3 (1990).

\bibitem{weinberg}
S.~Weinberg, Physica {\bf 96A},  327  (1979).

\bibitem{hbcpt}
E.~Jenkins and A.~V.~Manohar, Phys. Lett. {\bf {\bf B255}},  558  (1991).

\bibitem{weinberg2}
S.~Weinberg, Nucl. Phys. {\bf B363} 3 (1991).

\bibitem{we}
J.~W.~Bos, D.~Chang, S.~C.~Lee, Y.~C.~Lin, and H.~H.~Shih, Phys. Rev.
{\bf D51}, 6308  (1995).

\bibitem{dono}
J.~F.~Donoghue, E.~Golowich, and B.~R.~Holstein, {\em {Dynamics of 
the standard model}} ({Cambridge}, {New York}, 1992).

\bibitem{cole}
S.~Coleman and S.~L.~Glashow, Phys. Rev. Lett. {\bf 6},  423  (1961).

\bibitem{jenk92.2}
E.~Jenkins, Nucl. Phys. {\bf B375},  561  (1992).

\bibitem{lee-swift}
B.~W.~Lee and A.~R.~Swift, Phys. Rev. {\bf 136B}, 228 (1964).

\end{thebibliography}
\end{document}